\def\TMIT{$T_{\rm MIT}$}
\def\vo2{VO$_2$}
\def\alo{Al$_2$O$_3$}
\def\sio2{SiO$_2$/Si}
\def\cm{cm$^{-1}$}
\documentclass[aps,twocolumn,showpacs]{revtex4}
\usepackage{graphicx}%
\usepackage{graphicx}%
\usepackage{dcolumn}%
\usepackage{bm}%
\usepackage{xcolor}

\begin{document}
\title{Optical properties of VO$_2$ films at the phase transition: \\
influence of substrate and electronic correlations}
\author{Tobias Peterseim}
\affiliation{1. Physikalisches Institut, Universit\"at Stuttgart, Pfaffenwaldring 57, 70550 Stuttgart, Germany}
\author{Martin Dressel}
\affiliation{1. Physikalisches Institut, Universit\"at Stuttgart, Pfaffenwaldring 57, 70550 Stuttgart, Germany}
\author{Marc Dietrich}
\affiliation{I. Physikalisches Institut, Justus-Liebig Universit\"at Gie{\ss}en, Heinrich-Buff-Ring 16
35392 Gie{\ss}en, Germany}
\author{Angelika Polity}
\affiliation{I. Physikalisches Institut, Justus-Liebig Universit\"at Gie{\ss}en, Heinrich-Buff-Ring 16
35392 Gie{\ss}en, Germany}
\date{\today}

\begin{abstract}
Thin films of VO$_2$ on different substrates, Al$_2$O$_3$ and SiO$_2$/Si, have been prepared and characterized from room temperature up to 360~K. From the band structure in the rutile metallic phase and in the monoclinic insulating phase the optical properties are calculated and
compared with reflection measurements as a function of temperatures. Various interband transitions can be assigned and compared with previous speculations.
We extract the parameters of the metallic charge carriers that evolve upon crossing the insulator-to-metal phase transition and find effects by the substrate. The effect of electronic correlations becomes obvious at the phase transition.
\end{abstract}

\maketitle

\section{Introduction}
The metal-insulator transitions in the various vanadium oxides V$_x$O$_y$ have attracted unimpaired attention since decades for fundamental reasons as well as for possible applications \cite{Imada98,Katzke03,Basov11,Yang11
}. Among them vanadium dioxide \vo2\ is of particular importance because the phase transition slightly above room temperature ($T_{\rm MIT}\approx 340$~K) can be easily accessed and --~besides single crystals~--  high-quality thin films are available.
A structural phase transition takes place, when going from the metallic high-temperature rutile phase to the monoclinic low-temperature phase \cite{Morin59,Goodenough71}. Here the material is insulating without antiferromagnetic order, but does reveal charge ordering of vanadium pairs along the $c$-axis.
There has been considerable controversy over the relative importance
of a Peierls scenario with strong electron-phonon coupling
and electronic correlations representing Mott physics; nevertheless consensus has been reached that electron-electron interactions are an important aspect. Another interesting issue is the relation between the structural phase transition in \vo2\ and the first-order metal-insulator transition observed in various electronic properties
\cite{Zylbersztejn75,Wentzcovitch94,Rice94,Biermann05,Haverkort05,Qazilbash06,Kim06,Hilton07,
Arcangeletti07,Qazilbash09,Perucchi09,Ruzmetov10,Long12,Kang12}.

The transition can be tuned by doping, pressure, electric field and photoexcitation;
in the case of thin films, the influence of film growth temperature, thickness, orientation, doping, etc. on the physical properties has been subject of recent studies \cite{Youn04,Ruzmetov07a,Ruzmetov07b,Peng13,Dietrich15,Zhang15}. Here we measured the infrared optical properties of \vo2\ films and compared them with numerical results from density functional theory. Our findings confirm the idea of orbital ordering 
and indicate the importance of correlation effects at the phase transition.
We also disclose, however, that the evolution of the optical properties at the phase transition depends on the underlying substrate.

\section{Experimental Details and Characterization}
Vanadium dioxide thin films were deposited by radio frequency sputtering
using a metallic V target in a reactive process with an Ar/O$_2$ gas mixture.
During deposition, the sputtering gas pressure was about $2\times 10^{-4}$~mbar and
the Ar/O$_2$ ratio was 2.8\%. The growth temperature at the substrate table
was kept at $650^{\circ}$C over the full deposition period of 70~min \cite{Dietrich15}.

\begin{figure}
	\centering
		\includegraphics[width=1\columnwidth]{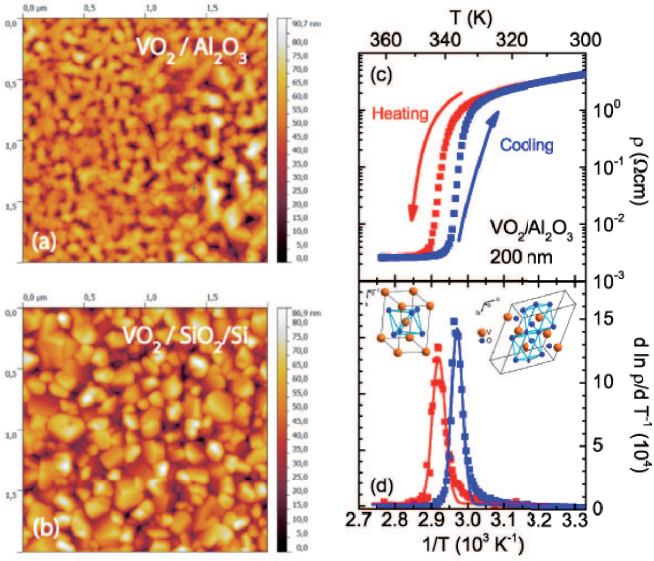}
	\caption{(a) and (b)~AFM pictures the VO$_2$ films on Al$_2$O$_3$ and Si substrates spanning an area of $2 \times 2\, \mu\text{m}^2$. The sapphire substrate is covered by VO$_2$-islands with an average diameter is $89 \pm 56\, \text{nm}$; in the case of the Si substrate the average grain size is  a factor of 1.5 larger.
(c)~Temperature-depend\-ent resistivity of VO$_2$ grown on sapphire in a $1/T$-plot. A clear hysteresis can be observed between the heating and cooling run. The resistivity drops/raises more than 3 orders of magnitude. The resistivity in the metallic state is almost constant whereas in the semiconductor state the resistivity exponentially increases on cooling following an activated behavior.
(d)~Plot of the derivative of log $\rho$ versus $1/T$. The maxima mark the phase transitions of the heating and cooling cycle. The transition temperature $T_{\rm MIT}$ for heating is 342~K whereas for the cooling the maximum is at $T_{\rm MIT}^{\rm cooling}=336$~K. In the semiconducting state the derivative is constant which implies an activated behavior with $\rho \propto \exp\left\{\Delta E/k_B T\right\}$ for $T<T_{\rm MIT}$.
The insets depict the crystal structure of VO$_2$ in the high-temperature rutile phase; the vanadium atoms are surrounded in octahedral manner by six oxygen atoms. In the monoclinic insulating phase the unit cell of VO$_2$ doubles.}
	\label{fig:AFM}	
    \label{fig:dc}
\end{figure}

The \vo2\ films with a thickness of approximately 200~nm were deposited on 1~mm thick sapphire ($c$-axis orientation) and silicon covered with a natural SiO$_2$ layer.  The final film thickness was measured by x-ray reflection  afterwards. The crystallite surface structure of the thin films was checked by scanning electron microscopy. The film quality was also monitored by atomic force microscopy; AFM pictures of the two samples on different substrates (roughness of about 1~nm) are displayed in Fig.~\ref{fig:AFM}(a) and (b). The average \vo2\ grain diameter  is $129\pm67$~nm on SiO$_2$/Si and $89\pm56$~nm on \alo;  this distinct difference will have severe influence on the physical properties as seen later.

The metal-insulator transition of the \vo2\ film on \alo\ is best seen in the temperature-dependent resistivity $\rho(T)$ plotted in Fig.~\ref{fig:dc}(c) as a function of inverse temperature. The large drop of $\rho(T)$ from $3.4~\Omega$cm in the insulating to $2.6\times 10^{-3}~\Omega$cm in the metallic phase demonstrates the high quality of the films. The thermal hysteresis extends from 336~K to 342~K as determined from a Gauss-fit of the derivative plotted in Fig.~\ref{fig:dc}(d). In the insulating side the conductivity follows an activated behavior with $\Delta E=(256\pm 4)$~meV; these values are in excellent agreement with previous studies \cite{Berglund69,
Ruzmetov07a}. With a mobility of $\mu_n=0.1$~cm$^{2}$/Vs we can determine the charge carrier density $N_c=5 \times10^{23}~{\rm cm}^{-3}$ in accordance with Hall measurements \cite{
Berglund69}. The rather large value implied that both electrons and holes contribute to the electronic properties.

The structural transition not only changes the space group from $P4_2/mnm$ to $P2_1/c$ \cite{Longo70,McWhan74}, it has severe implications for the electronic bands; hence we have calculated the bandstructure of \vo2\ in the rutile and monoclinic (M1) phase by density functional theory (DFT), as well as density of states and optical conductivity using norm-conserving general gradient approximate \cite{Perdew96
}; this goes beyond the local density approximation that is known not to describe \vo2\ properly \cite{Eyert02}. In the monoclinic phase 
a band gap of 0.68~eV occurs, in agreement with the experimental results \cite{Verleur68,Qazilbash06,Koethe06}.
\begin{figure}
	\centering
		\includegraphics[width=0.9\columnwidth]{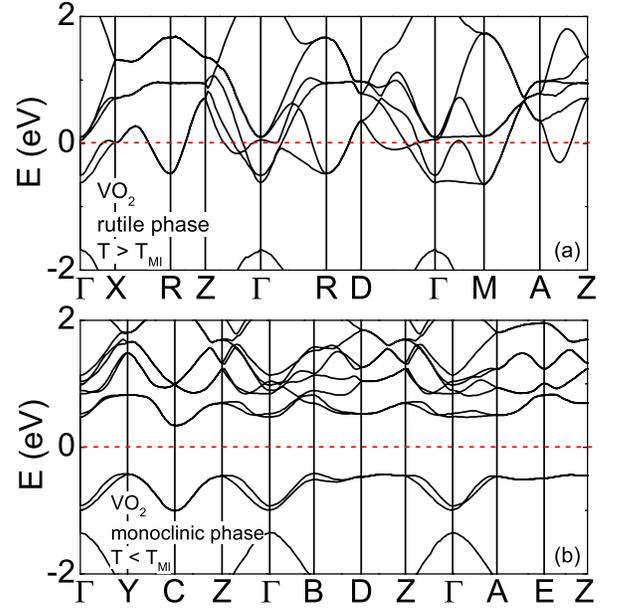}
	\caption{Band structure of VO$_2$ in (a) the rutile metallic phase with a space group $P4_2/mnm$ and (b) the  monoclinic M1 phase exhibiting a $P2_1/c$ crystal structure. The band gap is about $\Delta E = 0.68$~eV. }
	\label{fig:Bandstructure}	
\end{figure}

We have measured the optical reflectivity of the \vo2\ thin films on the different substrates as a function of temperatures in the mid-infrared range by employing a Fourier-transform infrared spectrometer Bruker Vertex 80v equipped with a hot-plate that allows for heating up to 400~K. We checked possible thermal gradients and ensured that the temperature values given here correspond to the actual film temperature.
Since for a 200~nm film the substrate properties affect the optical response, we cannot perform a simple Kramers-Kronig analysis to extract the intrinsic optical properties of \vo2.
Instead the optical conductivity of VO$_2$ was determined from the measured reflectivity spectra
via the Fresnel equations taking into account the known optical parameters of the substrate materials as well as multi-reflection \cite{DresselGruner02}. At each interface $ij$ the complex reflection and transmission coefficients are given by:
\begin{equation}
\hat{r}_{ij}=\frac{\hat{N}_i-\hat{N}_j}{\hat{N}_i+\hat{N}_j}\hspace*{1cm}
\hat{t}_{ij}=\frac{2\hat{N}_j}{\hat{N}_i+\hat{N}_j} \quad ,
\label{eq:Refl_Mulitsystem}
\end{equation}
where $\hat{N}_i=n_i+{\rm i}k_i$ denotes a complex refractive index.
In our case we have a two-layer system --~film on substrate~-- embedded in vacuum; hence the complex reflection coefficient $\hat{r}_{1234}$:
\begin{eqnarray}
&&\hat{r}_{1234}=\\
\label{eq:2Layer_Refl}
&&\frac{\hat{r}_{12}+\hat{r}_{23}\cdot e^{2{\rm i}\phi_2}+\hat{r}_{34}\cdot e^{2{\rm i}(\phi_2 + \phi_3)}+\hat{r}_{12}\hat{r}_{23}\hat{r}_{34}\cdot e^{2{\rm i}\phi_3}}{1+\hat{r}_{12}\hat{r}_{23}\cdot e^{2{\rm i}\phi_2}+\hat{r}_{23}\hat{r}_{34}\cdot e^{2{\rm i}\phi_3}+\hat{r}_{12}\hat{r}_{34}\cdot e^{2{\rm i}(\phi_2 + \phi_3)}} \nonumber
\end{eqnarray}
with $\phi_i=2\pi d \frac{(n_i+{\rm i} k_i)}{\lambda}$. This complex angle depends on the layer thickness $d$, the wavelength $\lambda$, the refractive index $n_i$ and the absorption index $k_i$ of the layer material. The power reflection $R$ is derived from $R=\sqrt{|\hat{r}_{1234}|^2}$. The optical properties of \vo2\ are described by a Drude term for the metallic charge carriers:
\begin{equation}
\hat{\sigma}(\omega)=\frac{N_c e^2\tau}{m}\frac{1}{(1-{\rm i}\omega \tau)} =\frac{\sigma_{dc}}{(1-{\rm i}\omega \tau)}
\label{eq:Drude}
\end{equation}
and the sum of Lorentzians
\begin{equation}
\hat{\sigma}(\omega)=\sigma_1(\omega)+{\rm i}\sigma_2(\omega)=\frac{N_c e^2\tau}{m}\frac{\omega}{(\omega + {\rm i} (\omega_0^2-\omega^2) \tau)}
\label{eq:Lorentz}
\end{equation}
for the phonons and interband transitions. Here $N_c$ is the density of contributing charge carriers,
$m$ their mass, $e$ the elementary charge, $\tau$ the scattering time, and $\omega_0$ the resonance frequency of the transition.

\section{Results and Discussion}

\begin{figure}
	\centering
		\includegraphics[width=0.9\columnwidth]{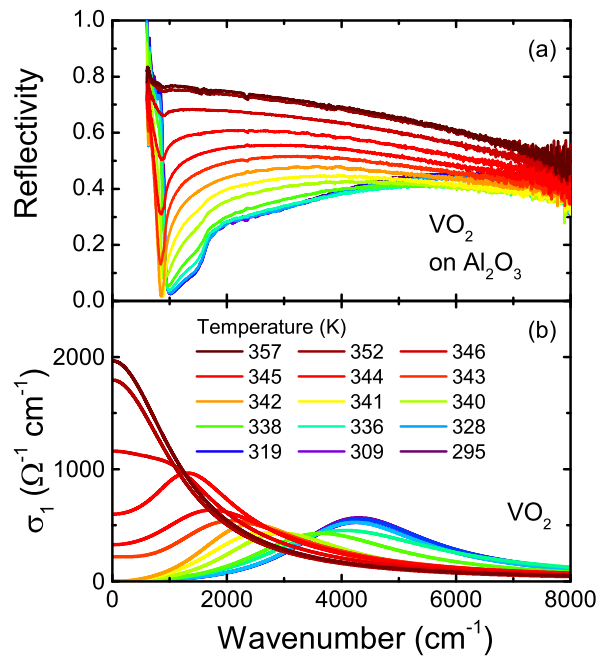}
	\caption{(a) Temperature-dependent and non-polarized infrared spectra of VO$_2$ on a sapphire substrate.  In the insulating state the reflectivity is on average 40 \% above 2000~\cm\ and almost transparent at 1000~\cm. Lattice vibrations of Al$_2$O$_3$ appear  at lower frequencies. With increasing temperature and crossing the phase transition the total signal rises up to $R=0.75$. Correspondingly the substrate features become suppressed.
(b)~Temperature-dependent optical conductivity of VO$_2$ determined by fitting the experimental data according to a two-layer model taking into account the substrate properties. In the insulating phase the spectrum of VO$_2$ consists of one interband transition reflected by the Lorentz peak. In the metallic phase the optical response consists solely of a Drude-feature.}
	\label{fig:VO2onAl2O3}	
\end{figure}

Temperature-dependent infrared measurement were performed for samples on both substrates SiO$_2$/Si and Al$_2$O$_3$ between 295 and 360~K in a frequency range from 800~\cm\ to 8000~\cm. In Fig.~\ref{fig:VO2onAl2O3}(a) the reflectivity of the VO$_2$-film on sapphire is illustrated for various temperatures. Between room temperature and 336~K, the reflectivity curves overlap and basically do not show any temperature dependence; for more elevated temperatures, the reflectivity strongly increases with heating.

\begin{figure}
	\centering
		\includegraphics[width=1\columnwidth]{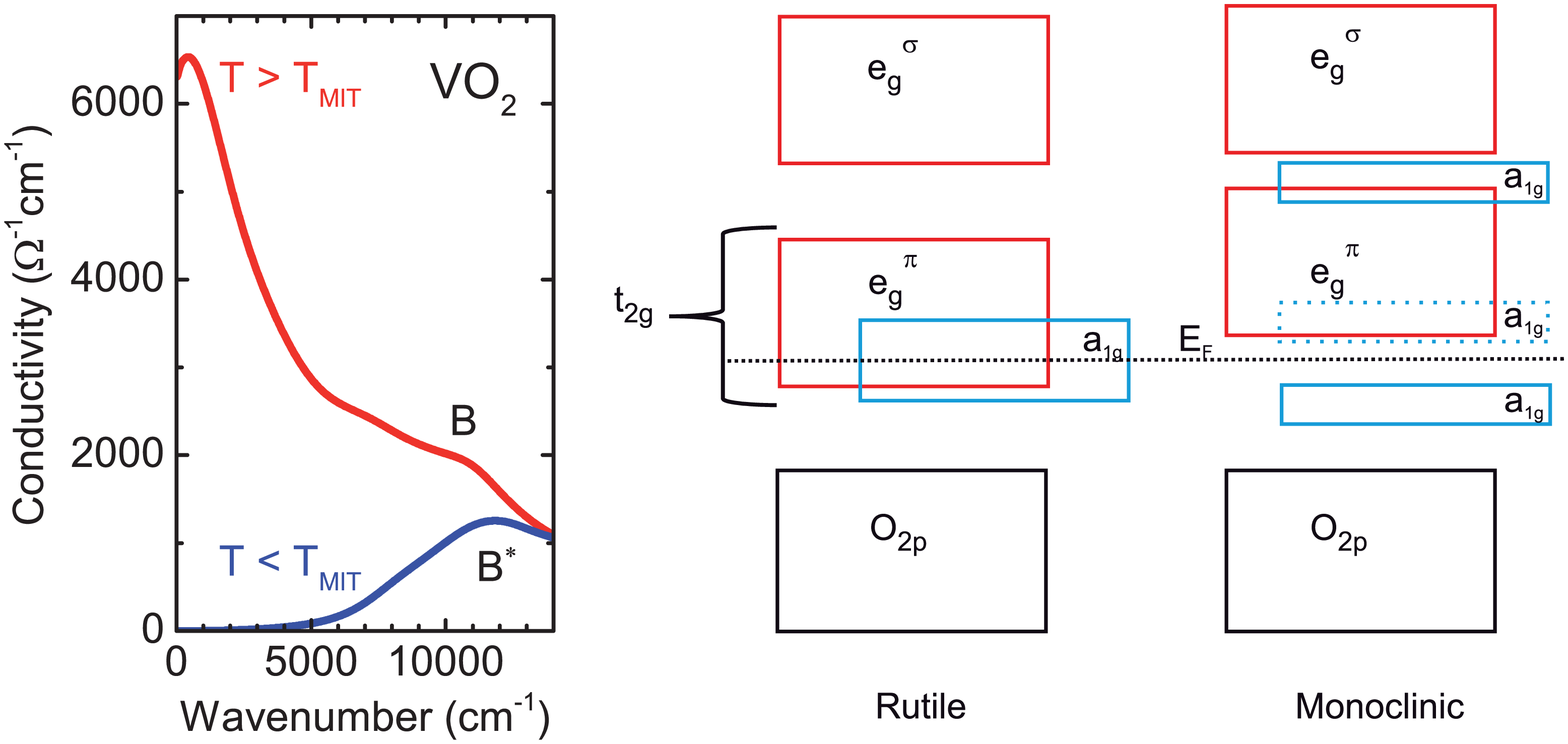}
	\caption{(left panel) Infrared spectra of VO$_2$ calculated from the electronic bandstructure
at 360~K (metallic rutile phase) and 295~K (insulating monoclinic phase) deplicted in Fig.~\ref{fig:Bandstructure}.
The simulated spectra reflect very well the evolution of the experimental data presented in Figs.~\ref{fig:VO2onAl2O3} and \ref{fig:VO2onSi}.
(left panels) Schematic bands of the rutile and monoclinic phase of VO$_2$. In the rutile phase the $t_{2g}$-band intersects the Fermi energy while it splits in the monoclinic phase due to the shift of the vanadium atoms in the octahedral cage and hence, an energy gap opens. Most likely the $a_{1g}$ consists of another part slightly above the Fermi level (dashed line).}
	\label{fig:cond}
	\label{fig:Bands}	
\end{figure}

First, the lattice phonon of the substrate should be noted at 800~\cm.
In the insulating state also an electronic interband transition of \vo2\ is present, which shifts to lower frequencies while approaching the phase transition. At room temperature it is located at 4300~\cm\ and moves to 1300~\cm\ upon heating to 343~K until it completely vanishes at higher temperatures.
Our bandstructure calculations presented in Fig.~\ref{fig:Bandstructure} revealed that VO$_2$ is a multiband system in which several valence bands contribute to the optical spectrum. In the energy range up to 1~eV (8060~\cm) the optical properties are dominated by the interband transition $a_{1g} \rightarrow e_{g}^{\pi}$ and $a_{1g} \rightarrow a_{1g}$. Qazilbash {\it et al.} concluded from their optical studies \cite{Qazilbash08} that the lowest transition takes places from $a_{1g}$ to $e_{g}^{\pi}$, but still higher than the one obtained from the fit to the VO$_2$/Al$_2$O$_3$ data.
However, photo\-emission experiments \cite{Koethe06} revealed that a fraction of the unoccupied $a_{1g}$ band is also located below the $e_{g}^{\pi}$-level --~but still above the Fermi energy~-- as sketched in Fig.~\ref{fig:Bands}, which was confirmed by calculations based on cluster dynamical mean field theory.
Accordingly, we suggest that the Lorentz function results from both, transition from $a_{1g}$ to $e_{g}^{\pi}$, as well as the transition from $a_{1g} \rightarrow a_{1g}$ which shift with the temperatures to lower energies.

Fig.~\ref{fig:cond} displays the theoretical conductivity of \vo2\ as calculated from the electronic density of states and bandstructure presented in Fig.~\ref{fig:Bandstructure}. In the metallic state $T_{\rm MIT}<T=360$~K a strong increase at low frequencies evidences the Drude-like behavior of the charge carriers. The shoulder B seen around 10\,000~\cm\ is in excellent agreement with the experimental spectra \cite{Verleur68,Qazilbash06}. It describes the interband transition from the partially filled $a_{1g}$ band into the unoccupied $e_{g}^{\pi}$-level.
The transition takes place between the $d_{yz}$-orbital (part of the $e_{g}^{\pi}$ band) of the neighboring vanadium atoms in the rutile plane and the $d_{x^2-y^2}$-orbital.

\begin{figure}
	\centering
		\includegraphics[width=0.9\columnwidth]{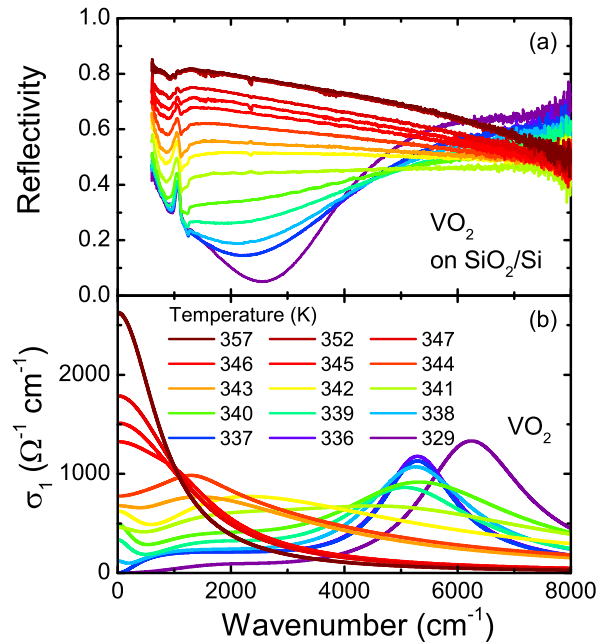}
	\caption{(a) Mid-infrared reflectivity spectra of VO$_2$ on SiO$_2$/Si in a temperature range from 330 to 360~K. The spectral shape deviates from the previous spectra of VO$_2$ on sapphire.
For the optical properties of the SiO$_2$/Si substrate most important is  a phonon resonance
that appears at about 1000~\cm\ and which can be assigned to SiO$_2$. Above the transition temperature the spectra reveals the same high reflectivity as VO$_2$ on sapphire.
(b) Temperature-dependence of the optical conductivity of  VO$_2$ determined by fitting the reflectivity spectra. In the insulating phase the spectrum consists of two interband transitions depicted by two Lorentz peaks. In the metallic phase only one Drude component is present.}
	\label{fig:VO2onSi}	
\end{figure}

While the spectrum in the metallic phase is dominated by free charge carriers and the interband transitions are screened, the optical conductivity for $T<T_{\rm MIT}$
clearly exhibit a band gap (Fig.~\ref{fig:cond}) of about 0.68~eV in very good agreement with the experimental results. The discrepancy to the activation energy derived from transport measurements [Fig.~\ref{fig:dc}(c)] is due to the fact that $\rho(T)$ probes indirect transitions, and both electrons and holes contribute to the electronic transport, as already concluded from the estimation of the carrier density $N_c$ above and Hall measurements \cite{Berglund69}.
In the insulating phase the corresponding band (B$^{\star}$) consists of excitations from the occupied $a_{1g}$ to the unoccupied $a_{1g}$ and $e_{g}^{\pi}$.
When approaching the phase transition the energetically lowest band $a_{1g}$ and $e_{g}^{\pi}$ shifts to smaller frequencies and reduces the band gap.
Upon crossing the insulator-to-metal transition a Drude component develops filling up the energy gap between $a_{1g} \rightarrow a_{1g}$ transition leading to the pseudogap like feature.
Correspondingly the interband transition at about 5000~\cm\ is modified and shifts to lower energies with rising temperature as observed in Figs.~\ref{fig:VO2onAl2O3}(b) and \ref{fig:VO2onSi}(b).

To describe the reflectivity of the VO$_2$ film on SiO$_2$/Si (Fig.~\ref{fig:VO2onSi}) the lattice vibration of silicon dioxide are most relevant, modeled by two Lorentz functions at 590~\cm\ and 1054~\cm.  A broad, weak mode is located at 1500~\cm\ and a stronger one at 5300~\cm;
it shifts to low frequencies at the phase transition.
With increasing temperatures a Drude component shows up and the Lorentz peaks lose intensity until they eventually vanish in the metallic phase. Here no effect of the substrate is observed any more.

\begin{figure}
	\centering
		\includegraphics[width=0.9\columnwidth]{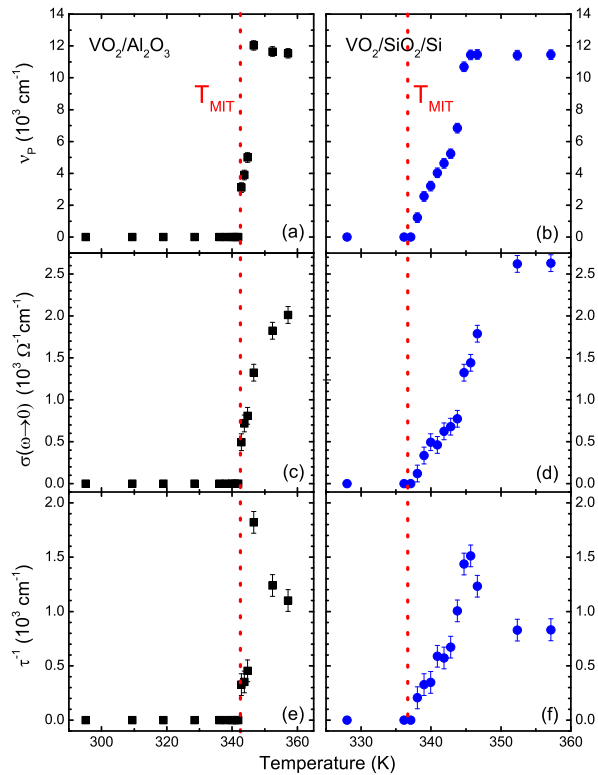}
	\caption{Temperature evolution of (a) the plasma frequency $\nu_{p}$,
(c) the conductivity in the static limit $\sigma_1(\omega\rightarrow 0)$  and (e) the scattering rate $\tau^{-1}$ of VO$_2$ on Al$_2$O$_3$ upon heating from $T=295$ to 360~K.  The panels (b), (d), and (f) on the right side display the corresponding parameters for VO$_2$ films on SiO$_2$/Si in a temperature range from 325~K to 360~K. It is to note that the right pictures show a less extended temperature range due to the higher temperature resolution. The red dashed lines mark \TMIT.}
	\label{fig:Parameters}	
\end{figure}

In Fig. \ref{fig:Parameters} the parameters of the Drude peak are shown for VO$_2$/Al$_2$O$_3$ as well as for VO$_2$/SiO$_2$/Si in the relevant temperature range around the respective metal-insulator transition. By the appearance of the Drude feature the transition temperature can be determined rather accurately: $T_{\rm MIT}^{\rm heating}=342$~K for sapphire and 336.5~K for SiO$_2$/Si. The transition temperature agreed exactly with the one obtained from transport measurements presented in Fig.~\ref{fig:dc}(c) and (d).
In both samples the plasma frequency $\nu_p= \omega_p/(2\pi c)$ increases strongly whilst crossing the phase transition and reaches a constant value of 12\,000~\cm\  in the metallic phase. $\omega^2_p =4\pi N_c e^2/m$ is a measure of the carrier density.
Similarly, the conductivity in the dc limit $\sigma(\omega\rightarrow 0)$ increases as well at the transition, flattens above 350~K in the metallic phase, and approaches asymptotically a constant value.
For SiO$_2$ it approaches $2000~\Omega^{-1}{\rm cm}^{-1}$, for Al$_2$O$_3$ a slightly larger value of
$2500~\Omega^{-1}{\rm cm}^{-1}$ in very good agreement with Ref.~\cite{Qazilbash06,Qazilbash07}.
The optical determined $\sigma_1(\omega\rightarrow 0)=\sigma_{dc}$ are by a factor 5 lower than the values derived from the transport measurements in Fig.~\ref{fig:dc}(c); in part this can be explained by the surface roughness. Interestingly, for the \vo2\ film on silicon the transition starts much more gradual at 337~K and exhibits a another step at $T_{\rm MIT}^{\rm heating}=344$~K where also a jump in the plasma frequency $\omega_p(T)$ is observed and where the scattering takes exhibits the divergency (see below). We relate this lower transition temperature to internal stress of the \vo2\ film on SiO$_2$/Si.

For both substrates the scattering rate $\tau^{-1}$ (Fig.~\ref{fig:Parameters} lower panels) reveals a non-monotonous behavior. At the phase transition it diverges and increases for the film on sapphire substrate to 1800~\cm\ and for SiO$_2$ to 1500~\cm; in the metallic phase it drops down by about 40\%\ to 1100~\cm\ and 800~\cm, respectively. The rise of the scattering rate marks the percolation boundary, meaning the transition from metallic islands to a closed metallic film. This behavior resembles
the divergency of the dielectric properties observed at the percolation threshold in ultrathin metal films \cite{Efros76,Tu03,Hovel10,DeZuani14}. Spatially resolved nano-spectroscopy reveals the growth of metallic islands, which eventually interconnect and percolate as the temperature crosses \TMIT.
The behavior of $\tau$ is also similar to temperature evolution of the effective mass at the phase transition \cite{Qazilbash07}.

With a Fermi velocity $v_F=0.24\cdot 10^{6}$~m/s (from DFT+LDA calculations \cite{Allen93}) the mean free path can be calculated from the estimated scattering rate;  at the transition we obtain $\ell = 7$~\AA\ and 20~\AA\ in the metallic phase. This supports a classical Boltzmann transport of the charge carriers. If $\ell \approx 10$~\AA\ or smaller the Boltzmann model breaks down. However, optical and transport measurements \cite{Allen93,Qazilbash06} reveal that in the case of VO$_2$ films $\ell$ is smaller than the lattice constant and the Ioffe-Regel-Mott limit violated. 

Performing our optical studies on films grown on two different substrates allows us to uncover the similarities as well as the difference; to the best of our knowledge no optical data were reported about VO$_2$ on SiO$_2$/Si. 
The spectra are plotted in Figs.~\ref{fig:VO2onAl2O3} and \ref{fig:VO2onSi}, the extracted optical parameters summarized in Fig.~\ref{fig:Parameters} for different temperatures.
In both compounds the phase transition from an insulating to a metallic state was observed very clearly in the spectra. The first interband transition in the low-temperature phase is located at 5300~\cm\ and lower in comparison to previous measurements. However, the band gap depends on the film quality and the amount of impurities leading to a slight spread of values. The jump in conductivity by three orders of magnitude is considered strong evidence of the excellent quality.
At the phase transition the scattering rate $\tau^{-1}$ diverges and drops in the metallic phase to 1100~\cm\ and to 800~\cm, respectively.

\section{Conclusions}
Thin films of VO$_2$ grown on Al$_2$O$_3$ as well as on SiO$_2$/Si substrates are extensively studied with different methods. AFM investigations shown in Fig.~\ref{fig:AFM} reveal an island-like growth for both substrates.~The diameter of the islands is about 100~nm on average exhibiting a height of 90~nm.
Thus the nominal film thickness of 200~nm results in an effective thickness of about 100~nm.
Measurements of the dc transport yield a conductivity jump of three orders of magnitude at the transition temperature $T_{\rm MIT}\approx$ 343~K. Also the width of the transition of about 6~K supports high quality of the films. The activation temperature $\Delta E=256$~meV and the charge carrier density $N_c=5\times 10^{23}~{\rm cm}^{-3}$ agree very well with literature values. It seems that the films on sapphire are of slightly superior quality compared to the one on silicon.

Theoretical calculations of the band structure based on the DFT-theory with a GGA-functional reproduce successfully that VO$_2$ is a metal in the rutile phase, but an insulator with an energy gap of about 0.65~eV in the monoclinic phase. The determined band structure supports Goodenough's model of the orbital ordering. The calculated optical spectra agree qualitatively with the experimental results and allow us to assign distinct features in the measured reflectivity  spectra to the corresponding calculated electronic excitations.

The optical spectra of the \vo2\ film were analyzed in the framework of a two layer-system. Depending on the electronic phase, Drude or Lorentz terms dominate. The extracted temperature-dependent parameters of the Drude feature enables us to trace the change of the different electronic parameters at the phase transition. The dc conductivity $\sigma_{1}(\omega\rightarrow 0)$ agrees well with previous measurements and transport data, but the plasma frequency $\omega_p$ seems to be somewhat low. Most interesting is the temperature behavior of the scattering rate $\tau^{-1}$ where a pronounced divergency at the phase transition was observed and a drop to half of its maximum value for elevated temperatures.
The divergent behavior is attributed to correlation effects at the phase transition.

\acknowledgements
We would like to thank E. Rose, D. Wu and S. Zapf for useful discussions as well
as G. Untereiner for technical support.
Funding by the Deutsche
Forschungsgemeinschaft (DFG) and the Carl-Zeiss Stiftung is acknowledged.


\end{document}